# Neutrinos from SN 1987A: Can they tell even more ?

H.-Thomas Janka[1]

Department of Astronomy and Astrophysics, University of Chicago,
5640 S. Ellis Avenue, Chicago, Illinois 60637, U.S.A.
Email: thj@mpa-garching.mpg.de

**Abstract**

The neutrino signal from SN 1987A provides an excellent opportunity to constrain physical theories for matter at extreme conditions and properties of particles that are produced in supernova (SN) cores. Phase transitions in the supranuclear equation of state (EOS) may change the cooling history of the collapsed stellar core and may even lead to a collapse of the protoneutron star to a black hole on a time scale of several seconds. The implications of such a scenario for the characteristics of the neutrino emission are tested. Moreover, the consequences of a reduction of the neutrino opacities by a possible suppression of the axial-vector interaction in a dense medium (Raffelt and Seckel 1994) are investigated. Since the strongest constraints from SN 1987A for the considered theories rely on the signal duration marked by the last three events in Kamiokande, we also discuss attempts to explain these events different from the standard protoneutron star cooling scenario. In particular, attention is focussed on the scenario of a rapid cooling of the protoneutron star within the time interval of the first 8 Kamiokande events or the IMB signal which is only about 2–5 s. Such a short neutrino emission could be the consequence of low neutrino opacities or convectively enhanced neutrino transport. Since the driving force of neutrino heating around the protoneutron star diminishes, the pressure support of the overlying material decreases and some matter might fall back to the neutron star. However, analytical estimates and numerical simulations indicate that the neutrino emission associated with the accretion of this material is hardly energetic enough to account for the last bunch of neutrinos at Kamiokande.

## 1. Introduction

Massive stars ($M \gtrsim 8\,M_\odot$) develop iron cores at their centers during a last hydrostatic burning phase that uses Si as fuel. When the mass of this core gets close to the Chandrasekhar mass, electron captures onto protons and nuclei and, for very massive stars, also photodisintegrations of nuclei speed up and reduce the pressure. Therefore the core becomes unstable against gravitational collapse. Within a fraction of a second the collapse proceeds to nuclear matter densities, where it is finally stopped due to the stiffening of the equation of state by repulsive nuclear forces and the pressure of non-relativistic (degenerate) nucleons. At this moment a strong shock wave is launched and starts to propagate outward in mass and radius. However, this shock is unable to cause the disruption of the progenitor star directly. Its propagation into the stellar mantle and envelope, which then leads to the spectacular outburst of light in a supernova explosion, is driven by the deposition of energy by neutrinos in the material behind the shock. These neutrinos are emitted in very large numbers from the hot protoneutron star that is formed at the center of the explosion and that releases its gravitational binding energy (several $10^{53}$ erg) in neutrinos.

---

[1] On leave from Max-Planck-Institut für Astrophysik, Karl-Schwarzschild-Str. 1, D-85740 Garching, Germany

On Feb. 23, 1987, at 7:35:35 UT($\pm 1$ min) a bunch of 11 neutrinos within a 12.4 s time interval was recorded by the Kamiokande II collaboration (Hirata et al. 1987) and another 8 neutrinos were counted in the IMB detector between 7:35:41.374 UT and 7:35:46.956 UT (Bionta et al. 1987). The Russian experiment at Baksan reported 5 events (Alexeyev et al. 1987) but has a rather large background rate and could uncover the connection with the other two detections only later due to problems with the time measurement. The neutrino events coincided in time with SN 1987A in the Large Magellanic Cloud, which was first observed only a few hours later. The overall characteristics of these neutrino detections like the event energies, the number of events, and the burst duration, agree well with theoretical predictions for the neutrino emission from stellar core collapse events. Both facts leave nearly no doubt that the neutrinos came from SN 1987A and thus mean the first detection of neutrinos from an extragalactic source.

This historical detection was considerably useful to constrain properties of neutrinos and of other particles that are possibly produced at the conditions in hot supernova cores. Moreover, the neutrino burst means a strong confirmation of principle aspects of supernova models and of the theoretical picture of the very complex processes that occur during supernova explosions. Certainly, the few neutrino events provide rather poor statistics and (therefore?) the recorded data show a number of unsatisfactory features, i.e., the weak compatibility of the Kamiokande and IMB signals, the clear peaking of the angular distributions away from SN 1987A, although isotropy was expected, and the 7 s gap between the eighth and the ninth event in Kamiokande. Nevertheless, the SN 1987A neutrinos have to be considered as a serious test for new physical models of matter at high densities or particle processes in a dense medium both of which may affect the evolution of supernovae and of new-born neutron stars. A nearby Galactic supernova will yield a much larger number of events. Hopefully, a more precise time measurement in different experiments and a much better temporal resolution will give us more detailed information about the spectral and temporal signature of the neutrino emission.

Although the neutrino signal yields the most direct constraint on the physical processes deep inside the exploding star, it is not the only one. Further constraints can be deduced from observed properties of neutron stars (masses, kick velocities, cooling behavior), from supernova nucleosynthesis, which was recognized during the past few years to depend sensitively on the character of the neutrino emission from the nascent neutron star, and from the explosion energy of a supernova, which is determined by the neutrino-matter interaction near the protoneutron star. Moreover, also the structure of the photon flux from a supernova provides information about the processes during the onset of the explosion. SN 1987A showed that the light curve is strongly affected by large-scale mixing which reaches deep into the star and which occurs already during the early moments of the explosion. Radioactive elements that are formed close to the center are therefore mixed into the He- and H-layers of the progenitor star with very high velocities and with significant anisotropy. This was deduced from infrared emission lines of Fe in the spectra of SN 1987A and from X-ray and $\gamma$-ray observations of SN 1987A and other supernovae. The variety of different observables which are influenced or determined by the physical events in the deep interior will make Type-II supernovae a very suitable high-energy physics laboratory once a standard model of supernova explosions has been developed.

In the presented paper the main results of two projects are summarized. In the first project we (Keil and Janka 1995) investigated the implications of hadronic phase transitions at supranuclear densities for the deleptonization and cooling of protoneutron stars. These phase transitions are a consequence of the production of new baryonic or mesonic states like hyperons, pions or kaons (Sect. 2). In the second project, described in Sect. 3, we (Keil, Janka, and Raffelt 1995) studied the consequences of a reduction of the neutrino opacities due to a possible suppression of the axial-vector currents in a dense medium (Raffelt and Seckel 1994; Janka, Keil, Raffelt, and Seckel, in preparation). In Sect. 4 some issues of non-standard models for the neutrino emission from supernovae will be discussed with respect to the question whether the 7s gap followed by the last three Kamiokande events might have a physical explanation rather than being considered as a statistical fluctuation.

## 2. Hadronic phase transitions at supranuclear densities

At high densities the Fermi momenta of nucleons can become sufficiently high to allow for the production of additional hadronic states besides $n$ and $p$. The physics at densities above nuclear matter density ($\rho_{\rm nuc} \approx 2.5 \cdot 10^{14}\,{\rm g/cm}^3$) is not well understood and a variety of suggestions for the occurrence of new particles can be found in the literature. Dependent on the assumptions about the nucleon-nucleon interaction the production of pions, associated with the existence of a $\pi$-condensate, has been considered (see, e.g., Takahara et al. 1991, Takatsuka et al. 1992), the formation of a condensate of $K^-$ was discussed (Brown et al. 1992, Mayle et al. 1993), and the effects of additional hyperonic states ($\Lambda$, $\Sigma$,...) or $\Delta$-resonances have been studied (e.g. Glendenning 1985, 1989).

Using finite temperature extensions of two $T = 0$ EOSs provided to us by N.K. Glendenning (personal communication) we investigated the implications of a hyperonization for the neutrino cooling of newly formed neutron stars. EOS A describes a gas consisting of $n$, $p$, $e^{\pm}$, and $\mu^-$, while EOS B contains hyperons and $\Delta$-resonances that are formed at $\rho \gtrsim 2\rho_{\rm nuc}$ via reactions like

$$N + N \longrightarrow N + \Lambda + K \qquad (1)$$

where the $K$ mesons subsequently decay into $\gamma$, $\mu$, and $\nu$. 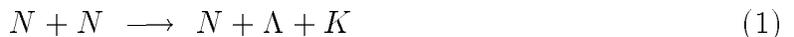 The occurrence of new hadronic degrees of freedom at high densities "softens" the EOS, i.e. flattens the pressure

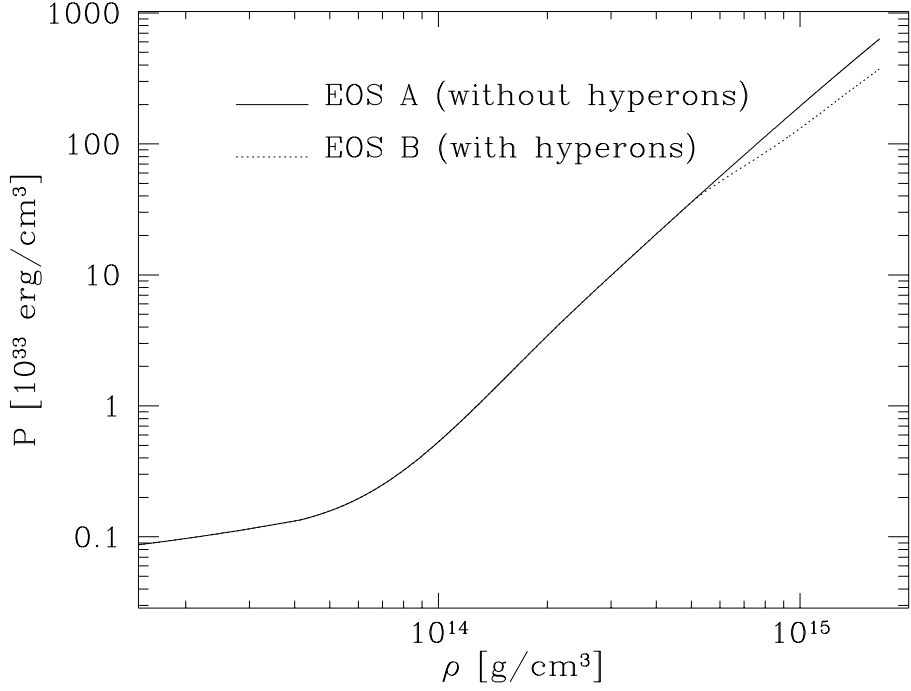

**Fig. 1.** Pressure as a function of baryon rest mass density, $\rho = n_b m_u$ ($m_u = 1.66 \cdot 10^{-24}$ g), for $\mu_{\nu_e} = 0$ and $T = 0.5\,\text{MeV}$.

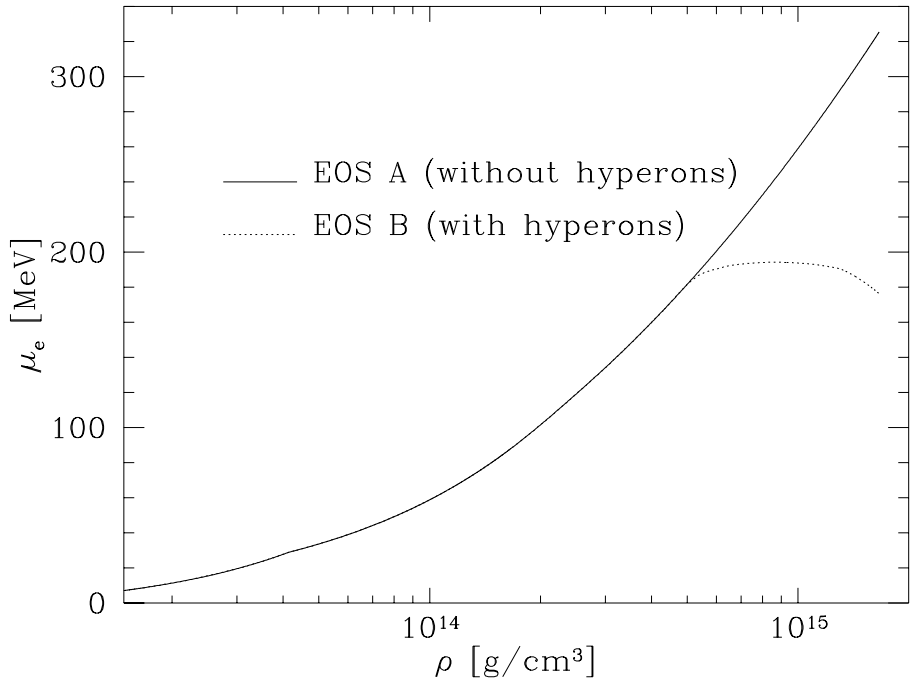

**Fig. 2.** Electron chemical potential as a function of baryon density $\rho$ for $\mu_{\nu_e} = 0$ and $T = 0.5\,\text{MeV}$.

increase $P(\rho)$ as a function of density $\rho$ (Fig. 1) because the nucleon degeneracy is reduced and degeneracy energy (or thermal energy), which is relevant for the pressure, is consumed in the production of massive particles. By reactions like those in Eq. (1) nucleons, in neutronized matter primarily neutrons, are converted into hyperons. This reduces the neutron chemical potential $\mu_n$ and leads to a maximum of the electron chemical potential $\mu_e = \mu_n - \mu_p$ of about 200 MeV in case of the hyperonic EOS B (for $\mu_{\nu_e} = 0$, i.e. deleptonized conditions) (Fig. 2). Although EOS B is a very special example and the inherent assumptions about physical parameters and processes may be largely different from the preferred choices and the use in other supranuclear EOSs, the behavior of $P(\rho)$ is qualitatively shared with a variety of EOSs (e.g., compare with Brown et al. 1992, Takahara et al. 1991, Takatsuka et al. 1992).

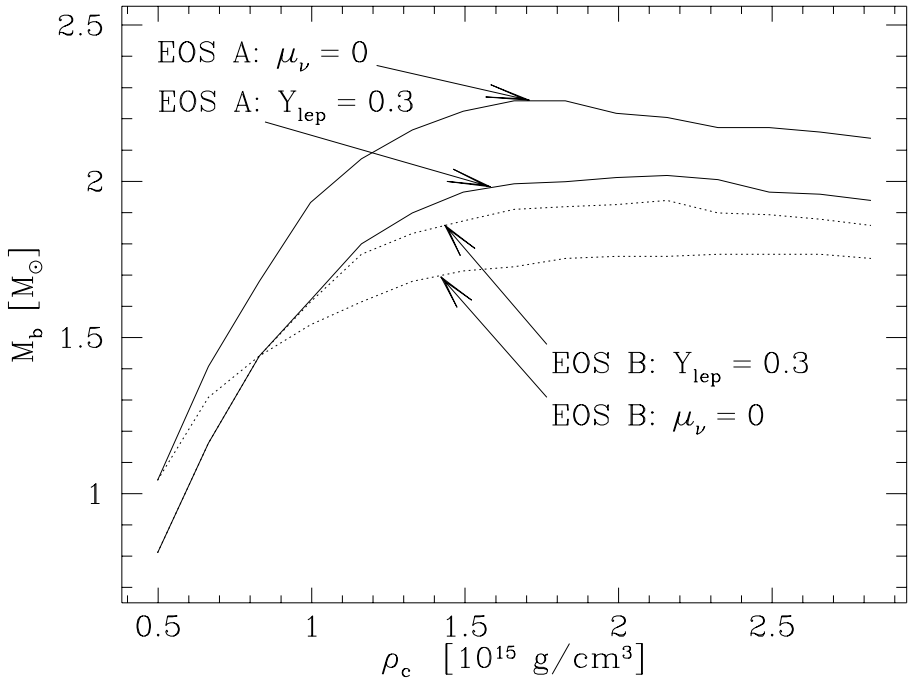

**Fig. 3.** Baryonic masses of neutron stars versus the central baryon rest mass densities $\rho_c$ of the stars. The two displayed cases for each EOS correspond to the lepton-rich ($Y_{lep} = 0.3$) and deleptonized ($\mu_{\nu_e} = 0$) situations. In all cases $T = 0.5$ MeV in the whole star was assumed.

The use of EOS B is also representative in its consequences for neutron star properties and the evolution of newly formed neutron stars. The effects of different EOSs can be visualized at their $M(\rho_c)$ relations which relate the neutron star masses to their central densities. Figure 3 displays the baryonic masses $M_b(\rho_c)$, defined as the total number of baryons $N_b$ times the atomic mass unit $m_u$, for the cold ($T \approx 0$) EOS A and EOS B both for the lepton-rich (lepton fraction $Y_{lep} = Y_e + Y_{\nu_e} = 0.3$) and deleptonized ($\mu_{\nu_e} = 0$) cases. The parts of the curves with negative gradients $dM_b/d\rho_c$ are unstable branches, i.e. the corresponding models are unstable against perturbations and cannot escape collapse on a dynamical time scale. Differences between the lepton-rich and deleptonized cases start to become significant for $\rho_c \gtrsim 5 \cdot 10^{14}\,\mathrm{g/cm^3}$. EOS A allows for neutron stars with larger maximum masses than EOS B. For EOS A the maximum baryonic mass is $2.02\,M_\odot$ (maximum gravitational mass $1.84\,M_\odot$) in the lepton-rich case and $2.26\,M_\odot$ ($1.96\,M_\odot$) for deleptonized configurations. EOS B has a limiting baryon mass of $1.94\,M_\odot$ ($1.77\,M_\odot$) for $Y_{\mathrm{lep}} = 0.3$ and $1.77\,M_\odot$ ($1.58\,M_\odot$) for $\mu_{\nu_e} = 0$. Although the creation of hyperons consumes energy and converts degeneracy or thermal energy into particle rest mass energy (thus moderating the pressure increase with density), the hyperonization usually is associated with a net energy release and a higher neutrino loss of the cooling protoneutron star compared to an equal-mass star with EOS A. This is, because the softer EOS leads to a more compact structure and a stronger gravitational binding.

It is particularly interesting to compare the evolution from the lepton-rich state of a collapsed stellar core to the deleptonized, final neutron star for EOS A and EOS B. The neutronization process, of course, conserves the total baryon number and therefore proceeds along a horizontal line in Fig. 3. Starting from a stable state for the lepton-rich EOS A such a track always leads to a stable deleptonized configuration, while for EOS B the situation can occur that the baryon mass of the collapsed stellar core is higher than the maximum stable mass for the deleptonized case. This will inevitably lead to a collapse of the cooling and neutronizing remnant to a more compact state, probably to a black hole. The collapse does not happen immediately but on a secular time scale of several seconds, because it is triggered by the gradual decrease of the lepton content of the neutrino-radiating, hot protoneutron star.

We evolved a set of protoneutron star models from the hot, lepton-rich state shortly after core collapse to the final, deleptonized, cold neutron star stage to determine the time scale until the collapse occurs and to investigate the corresponding neutrino signal (Keil and Janka 1995). The models had different masses and different initial temperatures and were constructed and evolved with either EOS A or EOS B. Model names beginning with "S" mean small baryonic mass, "L" denotes large mass (in case of EOS B above the maximum stable mass of cold neutron stars), "A" and "B" distinguish between use of EOS A or EOS B, and "C" and "H" indicate cool or hot initial models, respectively.

Figure 4 shows the central densities $\rho_c$ as a function of time. One can clearly see the difference of stable and unstable models. The central densities of the latter increase steeply when instability is approached while the stable models develop to finite values of $\rho_c$ for late times. All models reveal an initial drop of the density which is associated with a phase of increasing temperatures when the protoneutron star deleptonizes and degenerate electrons are converted into neutrinos which then downscatter and heat the stellar medium. Since the production of hyperons is favored as neutrons become more abundant in the star, instability occurs as a consequence of lepton number loss by $\nu_e$ emission. At the time of gravitational collapse the neutron stars are still hot. Accordingly, the stable models continue to radiate neutrinos with considerable fluxes for another ten seconds or even longer (Fig. 5). Since internal energy contributes to the gravitational potential and increases the star's gravitational mass, we could not find a significant stabilizing influence of thermal pressure once the leptons were gone and hyperonization had softened the supranuclear EOS. The additional mass that could typically be supported by thermal pressure was about 0.1–0.2 $M_\odot$. Once the unstable models start to contract and $\rho_c$ increases, their neutrino luminosities drop appreciably because of growing opacities and gravitational redshift. At the moment gravitational instability sets in and the increase of the central density accelerates, the neutrino emission is terminated and no late outburst of neutrino radiation occurs.

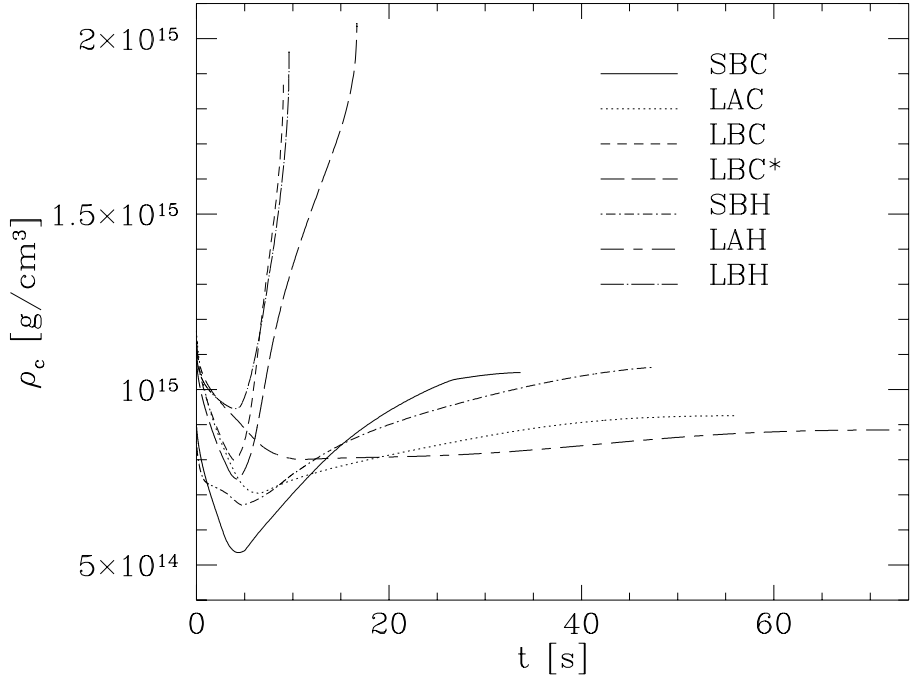

**Fig. 4.** Central baryon rest mass densities for the sample of computed protoneutron star cooling models versus (universal coordinate) time.

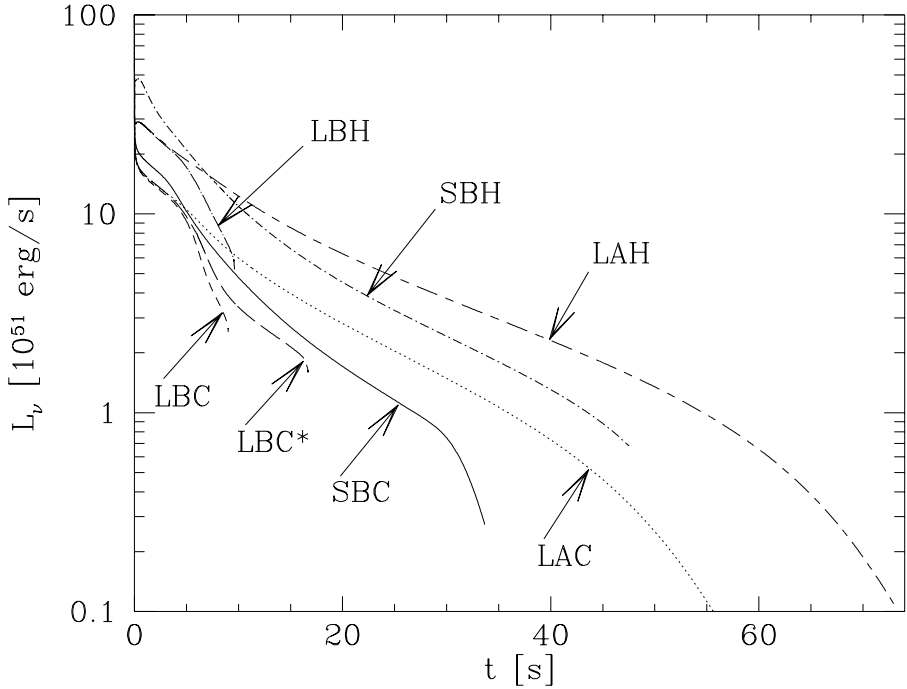

**Fig. 5.** Total neutrino luminosities for the sample of computed protoneutron star cooling models versus (universal coordinate) time. The luminosities are measured at the neutron star surfaces and include contributions from all neutrino types.

Typically, the collapse happens after less than 10 s of neutrino cooling. Model LBC* is an exception to this because both its baryonic mass (1.79 $M_\odot$) *and* its gravitational mass (1.67 $M_\odot$) were chosen only little above the maximum masses stabilized by EOS B for $\mu_{\nu_e} = 0$. For all other unstable models the gravitational masses were considerably higher above the mass limit of EOS B. Therefore only model LBC* could escape the onset of instability for nearly 17 s. Hot, collapsed stellar iron cores, however, have not yet emitted much of their gravitational binding energy in neutrinos and are usually weakly bound. This means that for "realistic" protoneutron star models the initial gravitational mass is not much different from the baryonic mass. The corresponding model LBH has a gravitational mass considerably in excess of the maximum stabilized mass of EOS B. Model LBC, which also collapses in less than 10 s, has a baryonic mass that is about 0.04 $M_\odot$ above the limit, in contrast to LBC* where it is only 0.02 $M_\odot$.

Since the deleptonization and gravitational instability of our models occurs in a time shorter than the signal duration at Kamiokande, we are able to draw the following conclusions. If the last 3 Kamiokande events between $t = 9.219$ s and $t = 12.439$ s (after the gap of about 7 s) are real events and do not have some non-standard explanation different from the commonly assumed protoneutron star cooling scenario, then either the mass of the neutron star in SN 1987A is below the limiting mass of EOS B ($M_b < 1.77\,M_\odot$) or a hyperonization and softening of the supranuclear EOS as described by EOS B is excluded on grounds of the detected neutrino signal. Of course, this only concerns the physical picture and the implications associated with EOS B (which is based on an effective relativistic field theory of interacting hadrons; Glendenning 1985) but does not mean that one has to discard the possibility of a formation of new hadronic states or even of hyperons in general. Also, we found that the collapse to a black hole happens "quietly" in the considered scenario of a continuous hyperonization, i.e. it is not accompanied by a final, energetic burst of neutrino radiation which might help to explain the peculiarities of the Kamiokande data.

## 3. Neutrino opacities and the SN 1987A neutrino signal

Nucleons in the neutron star medium are not isolated particles but interact with each other and collide frequently. A number of effects are discussed in the literature, for example many-body corrections like ion-ion correlations (Itoh 1975), nucleon correlations (Sawyer 1989) or Coulomb correlations of electrons and electron screening (Horowitz and Wehrberger 1991), which may lead to a reduction of the neutrino opacity in a dense medium compared to the opacity of a gas of independent nucleons. Recently, Raffelt and Seckel (1994) argued that the collisions among nucleons which interact via a spin-dependent force cause the nucleon spins to fluctuate fast relative to the other relevant time scales like the scattering time scale of neutrinos which is proportional to the inverse of the energy transfer. If the spin of a given nucleon flips several times during the characteristic interaction time, the effective nucleon spin "seen" by the neutrinos will be reduced by an averaging effect. Therefore, the axial-vector currents that mediate the spin-dependent part of neutrino-nucleon interactions might be nearly completely suppressed. Similar to Bremsstrahlung emission of neutrino pairs or axions this effect requires the presence of a bystander nucleon in a neutrino-nucleon scattering process, i.e. instead of the simple scattering $\nu + N \to \nu + N$ Raffelt and Seckel (1994) considered the reaction

$$\nu + N + N \longrightarrow \nu + N + N \ . \qquad (2)$$

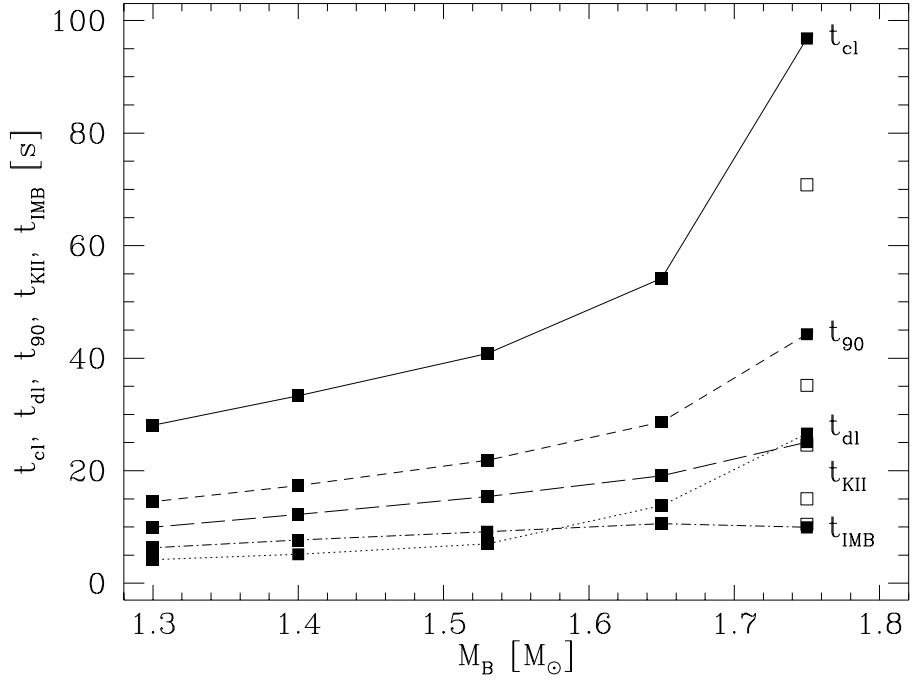

**Fig. 6.** Characteristic times of neutrino emission and expected signal durations in the KII and IMB detectors versus protoneutron star mass.

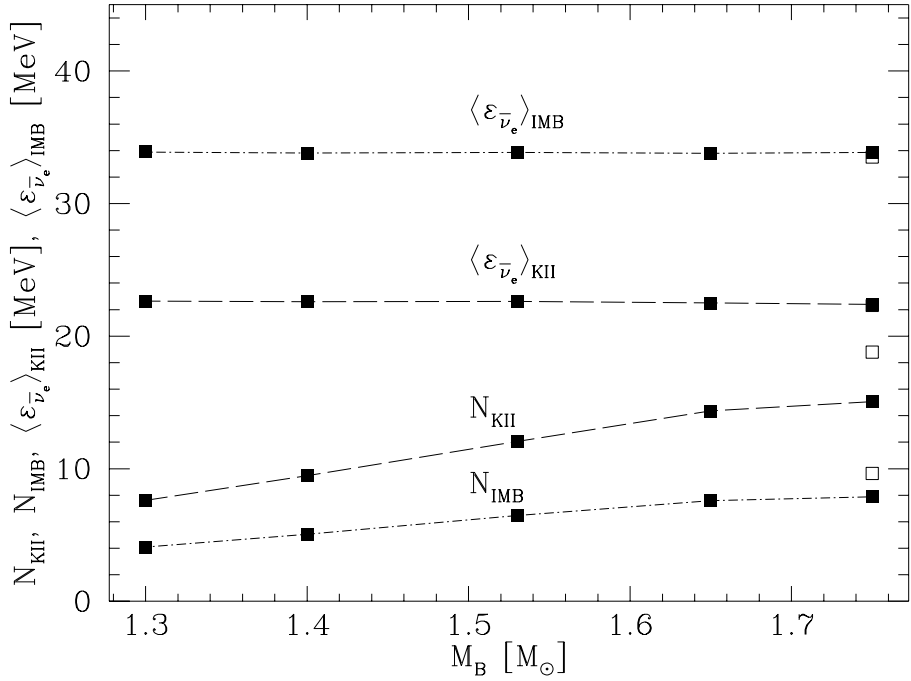

**Fig. 7.** Expected event numbers and mean energies of captured electron antineutrinos in the KII and IMB detectors versus protoneutron star mass.

Since axial-vector interactions with the axial-vector coupling constant $C_A$ yield the dominant contribution to the neutral-current neutrino-nucleon scattering cross section,

$$\sigma_{\nu N} \ = \ \frac{G_F^2}{4\pi} \left( C_V^2 + 3\, C_A^2 \right) \cdot \varepsilon_\nu^2 \ , \qquad (3)$$

(and are also the dominant part in charged-current $\beta$-processes) this would imply a significant reduction of the neutrino opacity at high densities. In Eq. (3) $C_V$ is the coupling constant for the vector current, $G_F$ the Fermi constant, and $\varepsilon_\nu$ the initial neutrino energy. The described effect adds to a possible suppression of the vector and axial-vector currents in a bulk nuclear medium, which leads to the reduction of the coefficients $C_V$ and $C_A$ relative to their vacuum values (see Raffelt and Seckel 1994).

In a pilot study we investigated the implications of the conjectured suppression of axial-vector currents for the cooling and deleptonization of newly formed neutron stars. We simulated the neutrino cooling for a large set of protoneutron star models with different masses, initial temperatures, EOSs (EOS A and EOS B), and neutrino opacities. The neutrino emission of these models was evaluated for the expected neutrino events in the Kamiokande II and IMB detectors. Demanding compatibility of the computed neutrino signals with the SN 1987A data allowed us to deduce limits on the strength of the suppression. For the purposes of this exploratory study it was sufficient to implement an opacity modification in a rather schematic way: we multiplied the standard axial-vector opacities with a suppression factor and combined the standard opacities with these suppressed opacities by weight factors $(1-a)$ and $a$, respectively, i.e. we replaced $C_A^2$ in Eq. (3) by

$$C_A^2 \times \left\{ (1-a) + a \cdot \left[ 1 + \left( \frac{\rho}{3 \cdot 10^{13}\ \mathrm{g/cm}^3} \right)^2 \left( \frac{10\, \mathrm{MeV}}{T} \right) \right]^{-1} \right\} \ . \qquad (4)$$

$a = 0$ corresponds to the standard case of no suppression, $a = 1$ yields full suppression, and with $a < 0$ we could also investigate the case of enhanced cross sections (for completeness).

Figure 6 shows the variation of characteristic times as a function of the neutron star mass (baryon mass). The standard values for neutrino opacities including fermion phase space blocking effects (see Keil and Janka 1995) were used in all models. Filled squares correspond to models computed with EOS B, open squares to those with EOS A. $t_{\rm dl}$ is the deleptonization time, $t_{\rm cl}$ the cooling time, $t_{90}$ is defined as the time interval after which 90% of the total energy emitted in neutrinos have left the star. $t_{\rm KII}$ denotes the time interval in which 90% of the predicted events in the Kamiokande II detector occur, and $t_{\rm IMB}$ gives the corresponding time for the IMB experiment. In Fig. 7 the average energies of captured $\bar{\nu}_e$ in both detectors, $\langle\varepsilon_{\bar{\nu}_e}\rangle_{\rm KII}$ and $\langle\varepsilon_{\bar{\nu}_e}\rangle_{\rm IMB}$, and the expected integrated numbers of events $N_{\rm KII}$ and $N_{\rm IMB}$ are displayed. These predictions have to be compared with the measured data of the SN 1987A neutrino signal: $t_{\rm KII}^{\rm 87A} \approx 12.4\,{\rm s}$, $t_{\rm IMB}^{\rm 87A} \approx 5.6\,{\rm s}$, $\langle\varepsilon_{\bar{\nu}_e}\rangle_{\rm KII}^{\rm 87A} \approx 17.4\,{\rm MeV}$, $\langle\varepsilon_{\bar{\nu}_e}\rangle_{\rm IMB}^{\rm 87A} \approx 33.9\,{\rm MeV}$, $N_{\rm KII}^{\rm 87A} = 11$, $N_{\rm IMB}^{\rm 87A} = 8$.

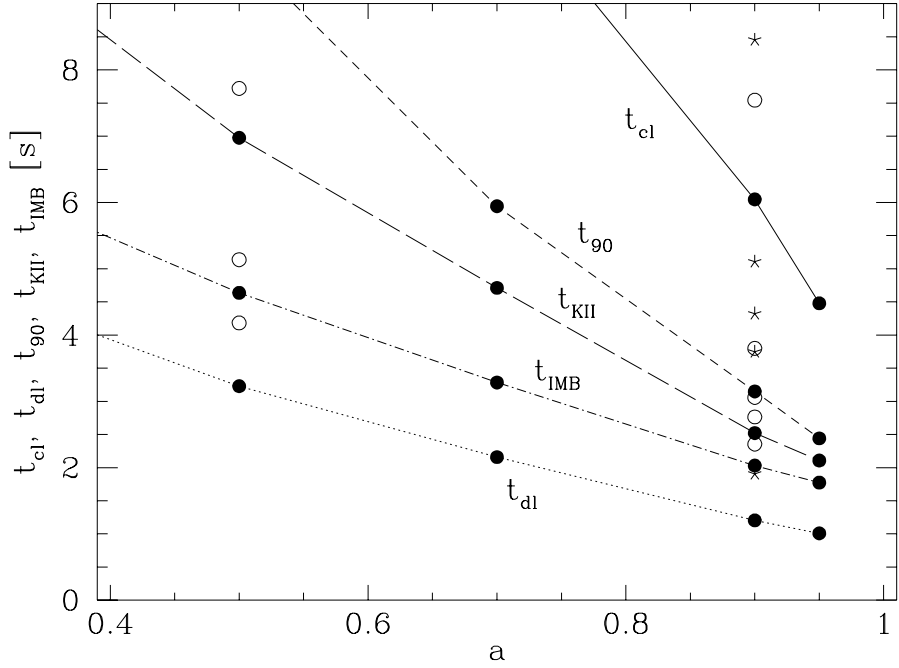

**Fig. 8.** Characteristic times of neutrino emission and expected signal durations in the KII and IMB detectors versus the parameter $a$. $a = 0$ corresponds to the standard neutrino opacities (see Fig. 6) and $a = 1$ means full suppression.

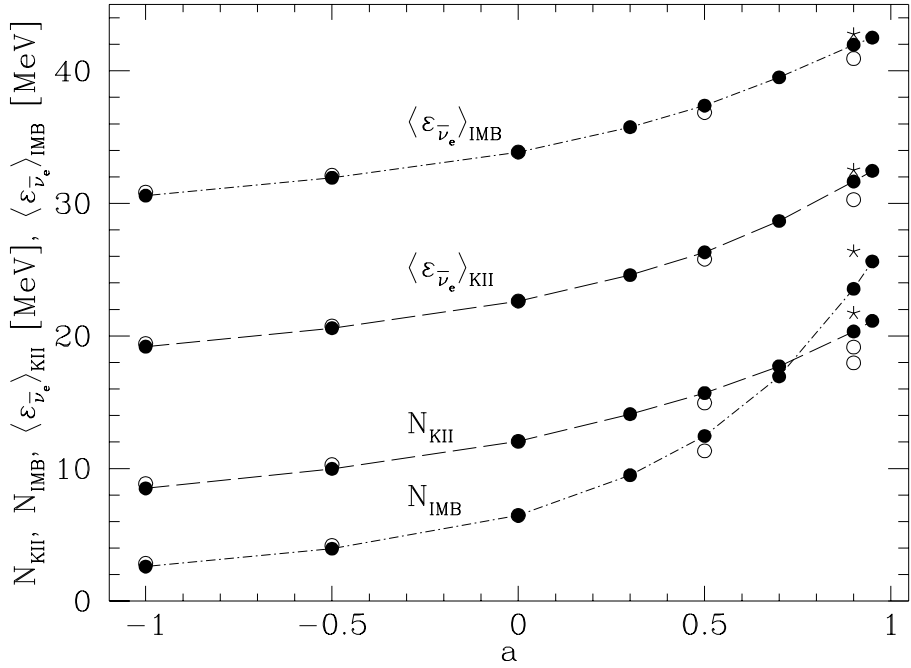

**Fig. 9.** Expected event numbers and mean energies of captured electron antineutrinos in the KII and IMB detectors versus the parameter $a$. $a = 0$ corresponds to the standard neutrino opacities, $a = 1$ means full suppression, and $a < 0$ belongs to the (hypothetical) case of enhanced opacities.

The average event energies in both detectors show remarkable stability, which probably reflects a kind of self-adjustment of the position of the neutrino sphere: higher temperatures in the neutron star (at a certain value of the density) cause higher opacities and therefore the neutrino sphere moves outward to a lower density where also the temperature is lower. In contrast, the characteristic times and event numbers increase with the neutron star mass and thus with the gravitational binding energy that is released in neutrinos. The flattening of $N_{\rm KII}$ and $N_{\rm IMB}$ and the slight decrease of $t_{\rm IMB}$ towards the right ends of the curves indicate the growing importance of gravitational redshift for massive neutron stars. This is particularly important for the IMB detector because of its high threshold energy of about 19 MeV and its slowly rising detection efficiency which make the detector very sensitive to the high-energy tail of the neutrino spectrum. Comparing the predicted signals with the SN 1987A data the best agreement is found for neutron stars with EOS B and a baryon mass between 1.5 and 1.6 $M_\odot$. Therefore we used our 1.53 $M_\odot$ model as a reference model for the calculations with varied neutrino opacities.

Figure 8 displays the variation of the times with increasing suppression of the neutrino opacities. Filled circles mark models where the axial-vector part in neutral-current scatterings on nucleons *and* in charged-current $\beta$-processes $\nu_e + n \leftrightarrow p + e^-$ and $\bar{\nu}_e + p \leftrightarrow n + e^+$ were changed in accordance with Eq. (4). Open circles indicate models where only neutral-current neutrino-nucleon interactions were modified, and asterisks denote the case where the critical density in Eq. (4) was taken as $10^{14}\,{\rm g/cm}^3$ instead of $3 \times 10^{13}\,{\rm g/cm}^3$. Figure 9 provides the results of expected event numbers and mean energies of captured neutrinos for the whole range of tested suppression parameters $a$. While all characteristic times drop drastically in case of reduced neutrino opacities, the event numbers and average event energies show the opposite tendency. Suppression of the neutrino opacities moves the neutrino sphere to higher densities and therefore higher temperatures. This "hardens" the emitted neutrino spectrum and leads to an increase of the predicted number of events. Note that due to the very steep density decline in the protoneutron star surface only very little matter is at densities below $3 \times 10^{13}\,{\rm g/cm}^3$ ($10^{14}\,{\rm g/cm}^3$) and the optical depth of this layer is rather small.

For $a \to 1$ both decreasing signal durations and increasing numbers and mean energies of the neutrino events lead to a clear discrepancy with the SN 1987A data. With a lower or higher protoneutron star mass or a different EOS not all aspects of this discrepancy can be compensated simultaneously. If the last 3 events in Kamiokande are interpreted as a part of the cooling signal of the nascent neutron star, we therefore con-clude that it is very unlikely that the neutrino opacity of the neutron star in SN 1987A was less than about 50% ($a \gtrsim 0.5$) of its standard value. From this limit on a possible suppression of axial-vector currents also a limit on the spin fluctuation rate in the nuclear medium can be deduced. The spin fluctuation rate seems to be of the order of the thermal frequency $\sim T$ and not as large as suggested by naive perturbative estimates which yield values of the order of 20–50 $T$ for typical conditions in protoneutron stars.

As discussed in detail by Keil, Janka, and Raffelt (1995) and by Janka, Keil, Raffelt, and Seckel (in preparation) the structural similarity of Bremsstrahlung emission of axions and neutrino-nucleon scattering in the presence of a bystander nucleon is expressed by the fact that the calculation of the rates of both processes involves the same spin-density structure function for the nucleons. The form of this structure function depends on the spin fluctuation rate. Upper limits on the spin fluctuation rate therefore translate into limits on a possible suppression of the axion emission rate as expected from the Landau-Pomeranchuk-Migdal effect (Raffelt and Seckel 1991). Due to the spin fluctuation rate being only of order $T$ the axion emission rate in a dense medium appears to be not as much reduced as would be the case for the naive perturbative estimate of the spin fluctuation rate. However, it is roughly a factor of 10 lower than assumed in previous models for the axion emission from supernova cores. Therefore the limits on the axion-nucleon Yukawa coupling are weakened and the bounds on the axion mass become about a factor of 10 less stringent, giving $m_a \lesssim 10^{-2}$ eV instead of the previously claimed limit of $m_a \lesssim 10^{-3}$ eV.

## 4. The last 3 Kamiokande events – evidence for a non-standard scenario ?

In the studies summarized in Sects. 2 and 3 the strongest constraints and limits were obtained from the duration of the SN 1987A neutrino signal. These constraints relied on the signal length in the Kamiokande detector as a measure of the time interval over which the major part of the neutrino emission of the nascent neutron star took place. Since the signal duration is determined by the last 3 Kamiokande events the strange time structure of the Kamiokande signal (as well as the marginal compatibility of the IMB and Kamiokande detections) requires a closer investigation of the question whether the gap of about 7 s and the following bunch of 3 events within 3.2 s indicate some non-standard scenario. If events 9–11 could be convincingly explained without considering them as part of the cooling emission of the protoneutron star the arguments used in Sects. 2 and 3 would be significantly weakened.

Suzuki (1989) performed a statistical analysis of the significance of the time structure of the Kamiokande signal. Assuming an exponential decay of the event rate in the detector he showed that for all choices of the decay time scale and initial event rate the probability of a 7 s gap followed by 3 events never gets higher than 1–2%. The maximum probabilities were obtained for rather long decay time scales, longer than about 15 s. Although being quite small these probabilities are not so small that they are imperative for a non-standard scenario, in particular not because IMB observed two events in the 7 s gap. It may also well be that a different model of the neutrino emission with a higher neutrino flux at early times and a less steep than exponential decline later increases the probability of the Kamiokande signal structure. Certainly, however, considerations of scenarios alternative to the simple "standard" protoneutron star cooling model are not unreasonable speculations.

In Sect. 2 we have already mentioned that phase transitions in the supranuclear EOS with a slow and gradual formation of new hadronic states, triggered by the progressing neutronization of the matter, are not able to increase the probability of the last 3 Kamiokande events because no late outburst of neutrino radiation occurs, even not in the case of a collapse to a black hole after about 10 s of protoneutron star cooling. But could accretion onto a black hole that forms at some time between 1.9 s (last event before Kamiokande gap) and 9.2 s (first event after gap) account for the bunch of the 3 late events? The delayed formation of the black hole could happen in a scenario similar to the one discussed in Sect. 2. Or could even fallback of matter to the neutron star yield sufficient neutrino emission to cause 3 events in the Kamiokande detector? In the latter scenario the protoneutron star might release its binding energy within a short time of only 2–6 s, the upper end of this time interval being set by events 7 and 8 in IMB. Such a short cooling time could mean a stronger suppression of the neutrino opacities than inferred from the analysis in Sect. 3. Due to the disregard of the last bunch of Kamiokande events and the correspondingly lower total energy required to explain the detections, this would also imply a somewhat smaller (initial) neutron star mass than estimated from the whole 12 s signal.

What are the implications concerning the IMB signal when the 7 s gap in the Kamiokande data is assumed to have a physical rather than statistical cause? Not dismissing the last two IMB events as background, the break in the neutrino detection of both experiments is actually only about 3.6 s. This is not much longer than the pause between the sixth (at 2.684 s) and seventh event (at 5.010 s) in IMB. Certainly, one has to ask whether it is imperative to search for alternative models of the neutrino emission from SN 1987A to explain a gap of only 3.6 s. Yet, the energies of the last two IMB counts (at 5.010 s and 5.582 s) are only slightly above the detector threshold ($19 \pm 5$ MeV and $22 \pm 5$ MeV). The low detection efficiency of IMB in this energy range requires a sizable neutrino flux which should have caused more than 100 neutrino events in the Kamiokande experiment that has a much lower threshold of about 7 MeV. The absence of a corresponding signal at Kamioka was interpreted as an indication that the last 2 IMB events might have occurred due to a statistical fluctuation or their energies were estimated systematically lower (Suzuki 1989). Dropping events 7 and 8 of IMB as background would certainly be a radical and rather unsatisfactory step. Keeping the last 2 IMB counts as real, however, lowers the need for a non-standard emission model and moreover will not allow one to improve the weak agreement between the Kamiokande and IMB measurements even with an alternative neutron star cooling model.

Can a non-zero neutrino mass and the corresponding flight delay of less energetic neutrinos account for the gap in the Kamiokande data (and the $\sim 2.7$ s break in IMB)? For two neutrinos of mass $m_\nu$ and energies $\varepsilon_1$ and $\varepsilon_2$ with $m_\nu c^2 \ll \varepsilon_1, \varepsilon_2$ one gets for the difference of the flight times

$$\Delta t_{12} \;=\; \frac{D}{2c}\left(m_\nu c^2\right)^2 \, \frac{\varepsilon_1^2 - \varepsilon_2^2}{\varepsilon_1^2 \, \varepsilon_2^2} \tag{5}$$

which is positive for $\varepsilon_1 > \varepsilon_2$. $D$ is the distance to the source. Plugging in representative numbers for the SN 1987A neutrinos, $\varepsilon_1 = 30\,\text{MeV}$ and $\varepsilon_2 = 10\,\text{MeV}$ and using $D \approx 50\,\text{kpc}$ as the distance to the Large Magellanic Cloud one obtains

$$\Delta t_{12} = 0.23\,\text{s}\,\frac{(m_\nu c^2)^2}{10\,\text{eV}^2}\,. \tag{6}$$

In order to explain a gap of several seconds between a bunch of energetic neutrinos that arrive earlier than a bunch of less energetic neutrinos a neutrino mass larger than 10 eV has to be assumed, which is excluded for electron neutrinos by current laboratory limits.

Let us therefore return to the accretion scenario already addressed above. The energy of a $\bar{\nu}_e$ burst that causes a number $N_\text{det}$ of events $\bar{\nu}_e + p \to e^+ + n$ in a water Cherenkov detector can be estimated to

$$E_{\bar{\nu}_e} \approx 1.7 \times 10^{51}\,\text{erg} \cdot \frac{1}{\mathcal{E}} \cdot \frac{N_\text{det}\,(D/50\,\text{kpc})^2}{(T_{\bar{\nu}_e}/5\,\text{MeV})\,(M_{\text{H}_2\text{O}}/2140\,\text{t})}\,. \tag{7}$$

$\mathcal{E}$ is the detector efficiency, averaged over the neutrino spectrum, $M_{\text{H}_2\text{O}}$ is the fiducial mass of the detector (2140 t in case of Kamiokande, 6800 t for IMB), and $T_{\bar{\nu}_e}$ is the spectral temperature of the $\bar{\nu}_e$ burst, the spectrum being described by a Fermi-Dirac distribution with zero degeneracy. The last $N_\text{det} = 3$ Kamiokande events have $T_{\bar{\nu}_e} = 1.6\,\text{MeV}$ (see Suzuki 1989). For $\mathcal{E} = 1.0$ one finds a lower limit on $E_{\bar{\nu}_e}$, $E_{\bar{\nu}_e} \gtrsim 1.6 \times 10^{52}$ erg. In a more careful evaluation Suzuki (1989) estimated $E_{\bar{\nu}_e} \approx 5^{+50}_{-4} \times 10^{52}$ erg. For the bunch of the first 8 events in Kamiokande ($T_{\bar{\nu}_e} \approx 3.3\,\text{MeV}$) one gets $E_{\bar{\nu}_e} \gtrsim 2 \times 10^{52}$ erg (with $\mathcal{E} = 1$), which is roughly the same number as inferred from the 8 IMB neutrinos, $E^\text{IMB}_{\bar{\nu}_e} \approx 2.4^{+8.2}_{-2.2} \times 10^{52}$ erg (see, e.g., Suzuki 1989, Janka and Hillebrandt 1989). Assuming equal energies emitted in all 6 kinds of neutrinos the total gravitational binding energy released by the collapsed stellar core should be $1\text{--}2 \times 10^{53}$ erg. This leads to a somewhat lower estimate of the (initial) mass of the neutron star in SN 1987A, for stiff EOSs between about $1.0\,M_\odot$ and $1.5\,M_\odot$, for soft EOSs only between $0.8\,M_\odot$ and $1.2\,M_\odot$.

In the accretion process onto a compact object with mass $M$, radius $R$, and redshift $e^{\Phi(R)}$ the accreted matter emits a fraction $f$ of its gravitational binding energy in neutrinos. The mass that has to fall onto the compact object to release an observable energy $E_{\bar{\nu}_e}$ in electron antineutrinos therefore is

$$\Delta M = (f f_{\bar{\nu}_e})^{-1} \cdot \frac{E_{\bar{\nu}_e}}{e^{\Phi(R)}\,GM/R} > 0.25\,M_\odot \left(\frac{E_{\bar{\nu}_e}}{5 \cdot 10^{52}\,\text{erg}}\right) \frac{R/10\,\text{km}}{e^{\Phi(R)}\,M/1.5\,M_\odot}\,. \tag{8}$$

The inequality neglects gravitational redshift and assumes an efficiency of conversion of gravitational binding energy into neutrinos of $f = 1$. The factor $f_{\bar{\nu}_e}$ is the fraction of the released energy that is radiated as $\bar{\nu}_e$, optimistically chosen to be $f_{\bar{\nu}_e} = 1/2$ (assuming that the energy is radiated only in $\nu_e$ and $\bar{\nu}_e$ and equally shared between both of them).

The mass that potentially falls back to the neutron star or black hole at the center of the supernova can roughly be estimated analytically, because at times later than about 2 s after shock formation a number of simplifying assumptions can be applied. From the virial theorem (Cooperstein et al. 1984, Bethe 1993a) the pressure behind the shock front can be estimated as

$$P_\mathrm{s} = \frac{E_\mathrm{sn}}{4\pi R_\mathrm{s}^3} \ . \tag{9}$$

$E_\mathrm{sn}$ is the supernova explosion energy, $R_\mathrm{s}$ the shock radius. The density outside the shock is given from the progenitor star by (Bethe 1990, Fig. 22)

$$\rho_0(r) = H\, r^{-3} \tag{10}$$

with $H \approx 6 \times 10^{31}$ g for an enclosed mass of $1.8\,M_\odot$ in the center of the $M \approx 18\text{--}20\,M_\odot$ progenitor of SN 1987A. From the Rankine-Hugoniot jump conditions in the shock one gets for the shock velocity

$$u_\mathrm{s} = \left(\frac{P_\mathrm{s}}{\rho_0(R_\mathrm{s})}\right)^{1/2} = \left(\frac{E_\mathrm{sn}}{4\pi H}\right)^{1/2} \approx 1.2 \cdot 10^9\,\mathrm{cm/s} \tag{11}$$

where the postshock pressure and density were assumed to be much larger than the preshock values, $P_\mathrm{s} \gg P_0(R_\mathrm{s})$ and $\rho_\mathrm{s} \gg \rho_0(R_\mathrm{s})$, and where the expression in the middle was obtained by using Eqs. (9) and (10). The numerical value gives the result for $E_\mathrm{sn} = 10^{51}$ erg. Note that $u_\mathrm{s}$ is independent of $R_\mathrm{s}$ which is not true in the start-up phase of the shock but is well fulfilled at later times when the shock is far away from the neutron star and pressure gradients as well as gravity have become small (see Bethe 1993b for an analytical discussion of this point). In that phase the kinetic energy (per unit mass) of the material dominates its gravitational binding energy and further acceleration of the matter behind the shock plays no role any more. The explosion energy of the supernova $E_\mathrm{sn}$ is then mostly kinetic energy. In this situation the separation radius $r_\mathrm{cut}$ of supernova ejecta and fallback material can be estimated by equating kinetic energy and gravitational binding energy,

$$\frac{1}{2}u^2(r_\mathrm{cut}) = \frac{G\,M}{r_\mathrm{cut}} \ . \tag{12}$$

Since according to Eq. (11) the shock velocity is constant, the velocity of the material at a point $r$ should be $\sim r/t$ where $t$ is the time after stellar core bounce and shock formation, thus

$$u(r_\mathrm{cut}) \sim \frac{r_\mathrm{cut}}{t_\mathrm{cut}} \approx r_\mathrm{cut}\frac{u_\mathrm{s}}{R_\mathrm{s}} \ . \tag{13}$$

For the same reason the shock position at the moment of separation, $t_\mathrm{cut} = t_0 + r_\mathrm{cut}/c_\mathrm{s}$, is approximately given by

$$R_\mathrm{s} \approx u_\mathrm{s}\,t_\mathrm{cut} = u_\mathrm{s}\,(t_0 + r_\mathrm{cut}/c_\mathrm{s}) \ . \tag{14}$$

$t_0$ is the time (after bounce) when the neutron star collapses to a black hole or when the neutrino luminosities decrease and the driving force by neutrino heating diminishes.

$r_{\rm cut}/c_{\rm s}$ gives a lower limit for the time in which the vanishing of the pressure support at the center is communicated outwards and arrives at the radius $r_{\rm cut}$ of the future mass cut. This information propagates outwards with a velocity that is somewhat smaller (delay by the inertia of the matter) than the sound speed $c_{\rm s}$ in the region behind the shock. Using Eq. (13) in Eq. (12) yields

$$\left(\frac{r_{\rm cut}}{R_{\rm s}}\right)^3 \approx \frac{2\,G\,M}{u_{\rm s}^2\,R_{\rm s}}\ . \tag{15}$$

Assuming constant density in the region enclosed by the shock, which is consistent with the assumption of a negligible pressure gradient, Eq. (15) gives an upper limit on the fraction of the shocked stellar material that falls back to the neutron star or black hole. The fallback mass is thus optimistically estimated as

$$\Delta M_{\rm acc} \approx \left(\frac{r_{\rm cut}}{R_{\rm s}}\right)^3 \cdot (M_{\rm s} - M) \tag{16}$$

when $M_{\rm s}$ is the mass enclosed by $R_{\rm s}$, i.e., the mass of the progenitor star which has been passed by the supernova shock by the time $t_{\rm cut}$. If the matter of the progenitor star ahead of the shock has not fallen inwards significantly since the onset of core collapse (which is well fulfilled at $t \gtrsim 2\,{\rm s}$ after bounce because the shock travels faster than the sound speed in the preshock material) then $M_{\rm s}$ can be calculated from Eq. (10):

$$M_{\rm s} \;=\; M_{\rm Fe} + \int_{R_{\rm Fe}}^{R_{\rm s}} {\rm d}r\, 4\pi r^2\, H\, r^{-3} \;\cong\; M_{\rm Fe} + 4\pi\, H\, \ln(R_{\rm s}/R_{\rm Fe})\ . \tag{17}$$

$R_{\rm Fe}$ is the radius of the stellar iron core before collapse, $M_{\rm Fe}$ the iron core mass, and $R_{\rm s}$ the shock position as given by Eq. (14). Combining Eqs. (16) and (17) and plugging in Eqs. (11), (14), and (15) our final result is

$$\Delta M_{\rm acc} \approx \frac{2\,G\,M}{[E_{\rm sn}/(4\pi H)]^{3/2}\,[t_0 + r_{\rm cut}/c_{\rm s}]} \times \left\{M_{\rm Fe} - M + 4\pi\,H\,\ln\left(\frac{\sqrt{E_{\rm sn}/(4\pi H)}}{R_{\rm Fe}}[t_0 + r_{\rm cut}/c_{\rm s}]\right)\right\}\ . \tag{18}$$

$\Delta M_{\rm acc}$ decreases with time, therefore early collapse to a black hole or rapidly decreasing neutrino fluxes and neutrino heating (i.e., small $t_0$) favor massive fallback. Since the matter behind the shock is radiation dominated (see Bethe 1993b for a discussion) and therefore the adiabatic index is $\gamma = 4/3$, one can easily estimate the sound speed $c_{\rm s}$ by again using the assumptions of constant pressure, Eq. (9), and constant density, $\rho_{\rm s} \approx (M_{\rm s} - M)/(4\pi R_{\rm s}^3/3)$:

$$c_{\rm s} \approx \left(\gamma\frac{P_{\rm s}}{\rho_{\rm s}}\right)^{1/2} \approx \left(\frac{4\,E_{\rm sn}}{9\,(M_{\rm s}-M)}\right)^{1/2}\ . \tag{19}$$

Typical masses $(M_\mathrm{s} - M)$ are 0.2–1 $M_\odot$ for the times of concern, yielding sound velocities of about $0.5$–$1 \times 10^9$ cm/s, which is only slightly less than the shock speed (Eq. (11)). In order to derive an upper limit of the accreted mass we therefore use the lower bound for the time in the denominator of Eq. (18), $t_0 + r_\mathrm{cut}/c_\mathrm{s} > t_0$, and replace the sound travel time $r_\mathrm{cut}/c_\mathrm{s}$ in the numerator by $t_0$ which certainly represents an optimistic case because $\ln(x)/x$ becomes maximum for $x = \mathrm{e} \approx 2.72$. Plugging representative numbers into Eq. (18), $M = 1.5\,M_\odot$, $E_\mathrm{sn} = 10^{51}$ erg, and $R_\mathrm{Fe} = 3000$ km, and taking a favorable, large value of the difference of stellar iron core mass and initial neutron star mass, $(M_\mathrm{Fe} - M) \approx 0.2\,M_\odot$, one ends up with a maximum mass accreted within a period of seconds of

$$\Delta M_\mathrm{acc} < 0.16\ M_\odot \qquad (20)$$

when the fallback is initiated at a time $t_0 = 2$ s.

Although the inequalities in Eq. (8) and Eq. (20) were derived on grounds of very optimistic assumptions, the fallback mass needed to account for the three late neutrino events in Kamiokande seems hard to reach for the typical conditions in the collapsed stellar core of an 18 $M_\odot$ star. Our estimate of $\Delta M_\mathrm{acc}$ depends on the quantity $H$ which varies with the mass of the progenitor star. For more massive stars $H$ is larger. Yet, the estimate of Eq. (18) was obtained by neglecting pressure gradients and, in particular, by assuming constant density inside the shock radius $R_\mathrm{s}$. This certainly overestimates the mass close to the newly formed neutron star and thus the fallback. Numerical simulations show that the ejected mass is moving in a thick, dense shell behind the outward propagating shock while the neutron star is surrounded by a low-density, high-entropy region ("hot bubble"). Accordingly, the fallback turns out to be more than an order of magnitude smaller than the limit of Eq. (20) in numerical models that include the formation of the hot bubble and the neutrino wind from the protoneutron star (Keil and Janka, unpublished). Interestingly, the mass accretion rate in these models peaks at about 10 s after core bounce and shock formation.

It must be emphasized once again that both the lower bound on the required accretion mass, Eq. (8), *and* the upper limit on the possibly accreted mass, Eq. (20), were obtained by using extreme assumptions throughout. The numerical value in Eq. (8) assumes that gravitational redshift is negligible, that the conversion efficiency of gravitational binding energy into neutrinos is 100%, and that only $\nu_e$ and $\bar\nu_e$ are emitted and carry away equal amounts of energy. Equation (20) on the other hand is too optimistic because $t_0 = 2$ s is the absolute minimum for the time when the re-collapse of some matter might have been initiated in SN 1987A, and, moreover, because the density and pressure gradients between neutron star and supernova shock and the outward push by the neutrino wind were disregarded. Not even for all these extreme assumptions do the results of Eqs. (8) and (20) get close.

However, the analytical considerations in this paper and the numerical studies by Keil and Janka are based on very simplified models. They do so far neither include effects due to a reverse shock (which can already form at about 0.5 s after bounce, see Janka and Müller 1995, and might increase the fallback on a time scale of several seconds) nor do they take into account multi-dimensional effects like rotation or turbulence. Therefore the question of fallback masses and time scales requires further investigations.

# 5. Discussion and conclusions

Two exemplary studies were reported where the neutrino signal from SN 1987A was used to constrain properties of the supranuclear EOS and of neutrino-matter interactions. In particular, the duration of the neutrino pulse in the Kamiokande experiment marks a minimum duration of the neutrino emission which allows one to exclude scenarios which predict a more rapid cooling of the nascent neutron star. Such conclusions rely on the correct interpretation of the late 3 events in Kamiokande as a part of the protoneutron star cooling emission. Moreover, considering and normalizing changes due to the introduction of new physics relative to the results of a "standard cooling model" requires the acceptance of the standard as a meaningful model of SN 1987A. The standard neutrino physics in our models is based on the description of the interaction targets as particles of nondegenerate or degenerate, ideal Fermigases at thermal equilibrium, an approach which is generally used in supernova modelling. Phase space blocking effects are included in the formulation of the neutrino-matter interactions (Burrows and Lattimer 1986, Keil and Janka 1995).

Our standard treatment yields very good agreement with neutron star models calculated by other groups (Burrows and Lattimer 1986, Wilson and Mayle 1989, Suzuki 1989) although the numerical description employed by those groups partly differed significantly from the one chosen in our models. Moreover, the predicted neutrino signals from models with standard physics are in satisfactory agreement with the neutrino pulse from SN 1987A. This is quite astonishing, considered the facts that the neutrino-matter interactions in dense matter are poorly understood and there are a number of effects that should reduce the cross sections significantly compared to their naive values (e.g., particle correlations and screening; spin fluctuations as discussed in this work; smaller coupling constants in the bulk nuclear medium and relativistic corrections both for nucleon recoil and form factors; Seckel, personal communication). Moreover, the protoneutron star cooling models are spherically symmetrical, thus do not regard possible convective processes which might also accelerate the neutrino loss. The long duration of the neutrino emission from SN 1987A indicates that the protoneutron star is rather opaque to neutrinos but the current knowledge of the source of this opacity is very unsatisfactory.

In detail our results were the following:
(i) Formation of hyperons and the associated softening of the supranuclear EOS can lead to a delayed collapse of the newly formed neutron star to a black hole if the neutron star mass is above the maximum stable mass of the cold, deleptonized EOS. Since the instability is primarily triggered by the progressing deleptonization of the star, the collapse occurs on the neutrino diffusion time scale which is of the order of several seconds. Since we found the lepton-loss time to be shorter than the signal duration at Kamiokande we concluded that either the neutron star in SN 1987A was below the mass limit of the employed EOS, $1.58\,M_\odot$ (gravitational mass, or $1.77\,M_\odot$ baryonic mass), or that the suggested phase transition and softening of the EOS due to the creation of hyperons does not occur. Last but not least we found the onset of gravitational instability to happen "quietly" i.e. without a late, luminous outburst of neutrino radiation.
(ii) The duration of the neutrino pulse from SN 1987A and, to a minor degree, also the average event energies and event numbers in the Kamiokande and IMB detectors allowed us also to constrain the magnitude of a possible suppression of the axial-vector interactions of neutrinos with nucleons. This suppression was conjectured by Raffelt and Seckel (1994) as a consequence of rapid spin fluctuations of the nucleons which are caused by frequent collisions in a dense medium and lead to an averaging effect and therefore to a reduction of the effective target spin "seen" by the neutrino. In contrast to naive estimates based on pertubation theory which suggest spin fluctuation rates of the order of 20–50 $T$ for typical conditions in a newly formed neutron star, we deduced from our cooling models that the spin fluctuation rate should be of order $T$ only. This corresponds to a moderate, global reduction of the neutrino opacity by roughly 50%. It also implies that the axion emission rate from a collapsed stellar core is considerably lower than previously used values. In particular, the bounds on the axion mass derived from SN 1987A are weakened by roughly a factor 10 to $m_a \lesssim 10^{-2}$ eV.
(iii) Since our conclusions rely on the duration of the neutrino emission from the protoneutron star as marked by the last 3 events at Kamiokande, one has to test possible alternative scenarios for their ability to explain these late counts and the preceding pause of several seconds in the neutrino detection. An obviously attractive scenario could be a fast decline of the neutrino luminosities from the protoneutron star in a time of 2–5 s and a subsequent "quiet" phase of low neutrino fluxes, followed by accretion of fallback material which possibly yields sufficient neutrino emission to cause the last bunch of Kamiokande events. The rapid cooling of the protoneutron star could be a consequence of neutrino opacities that are lower than their standard values or might be caused by convectively enhanced neutrino transport. Due to the low energies of the three late neutrinos they correspond to an appreciable neutrino flux and thus a sizable fallback mass is required. Analytical as well as numerical studies show, however, that even with very optimistic assumptions it seems very unlikely that in case of an 18 $M_\odot$ progenitor star of SN 1987A fallback occurs which is massive enough to match the considered scenario.

### Acknowledgements.

The author would like to thank W. Hillebrandt, W. Keil, and G. Raffelt for a reading of the manuscript and interesting comments. This work was supported by the Sonderforschungsbereich 375-95 für Astro-Teilchenphysik der Deutschen Forschungsgemeinschaft. The author also acknowledges support by the National Science Foundation under grant NSF AST 92-17969, by the National Aeronautics and Space Administration under grant NASA NAG 5-2081, and by an Otto Hahn Postdoctoral Scholarship of the Max-Planck-Society. The computations were performed on the CRAY-YMP 4/64 of the Rechenzentrum Garching.